# Geo-Encryption Protocol For Mobile Networks

Ala Al-Fuqaha, Omar Al-Ibrahim, Ammar Rayes[*]
Department of Computer Science
Western Michigan University
Kalamazoo, MI
{alfuqaha , oalibrah } @cs.wmich.edu

*Abstract*— We propose a geo-encryption protocol that allow mobile nodes to communicate securely by restricting the decryption of a message to a particular location and time period. Our protocol will handle the exchange of movement parameters, so that a sender is able to geo-encrypt messages to a moving decryption zone that contains a mobile node's estimated location. We also present methods for estimating the node's movement parameters to allow for geo-encryption. Finally, we evaluate our model by measuring the induced overhead to the network and its performance in terms of decryption ratio.

## I. INTRODUCTION

GPS-based encryption (or geo-encryption) is an innovative technique that uses GPS-technology to encode location information into the encryption keys to provide location based security. GPS-based encryption adds another layer of security on top of existing encryption methods by restricting the decryption of a message to a particular location and time period. Applying this technique to a mobile environment, with a dynamically changing topology, requires a protocol to handle the distribution of movement information so that communicating hosts can keep track of each others locations. Existing GPS-based encryption techniques ([3] and [4]) have limited support for mobile nodes, therefore, we propose a mobility model for existing geo-encryption techniques to allow mobile nodes to exchange movement parameters so that a sender is able to geo-encrypt messages to a moving decryption zone that contains a mobile node's estimated location. We also simulate this protocol to find out its performance and scalability in a multi-hop network.

The paper is organized as follows: Section II introduces the geo-encryption model. Section III proposes a mobility model by introducing some movement parameters and the protocol required to set up and maintain a mobile geo-encrypted session. Section IV presents the model equations. Section V shows how to estimate and update the mobility parameters and how to achieve the model's objectives. Sections VI – IX present the simulation details. Section X concludes the paper.

## II. DENNING'S MODEL OF GEO-ENCRYPTION

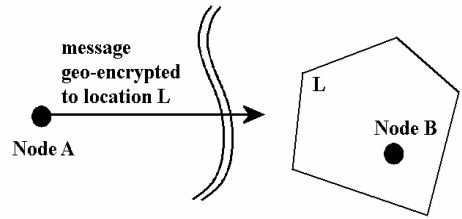

Figure 1: Geo-encryption forces the receiving node (B) to be inside a pre-defined location L to be able to decrypt the message.

Denning's model ([1], [2] and [3]) for adding security to transmissions uses location-based encryption to limit the area inside which the intended recipient can decrypt messages. Fig. 1 shows the general idea. A geo-locking function is employed during the encryption process to combine an encryption key with the recipient's geographic location (L) to produce a "geo-secured" key for transmission alongside an encrypted message; the message can only be decrypted if the geo-secured key can be recovered, which can only be done if the recovering machine is physically positioned at location L. The sender also transmits parameters which define the shape of the area where decryption is permitted (the "decryption zone"), and the time period during which decryption can be accomplished.

Denning's model is effective when the sender of a message knows the recipient's location L and the time that the recipient will be there, and can be applied especially effectively in situations where the recipient remains stationary in a well-known location. Denning recognizes that geo-encryption is also desirable in situations where the recipient is mobile, without a pre-planned itinerary. All the sender needs is the recipient's current/upcoming location and the time that the recipient will be there, and Denning notes that a velocity parameter (the recipient's velocity) can be added to the geo-locking function. However, we have not seen the details of mobility support in Denning's geo-encryption model, and therefore we propose a model to provide for mobility when using GPS-based encryption.

[*] Cisco Systems, Inc. E-mail: rayes@cisco.com

## III. THE MODEL

We propose a mobility model based on the geo-encryption technique in [1] in which both sender and receiver are mobile, without preplanned itineraries, and can securely deliver their current locations to one another whenever necessary. In order to do this, each mobile node that will be receiving geo-encrypted messages needs to inform potential sender nodes about its intended movement in order for a sender node to estimate the mobile node's expected location at any point in time. This is done by sending information regarding the mobile node's movement, which we call mobility parameters, to the sender via a sequence of message exchanges. The next two subsections explain these parameters, the protocols and the messages that need to be exchanged between a mobile receiver node and a stationary or a mobile sender node.

### A. Mobility parameters

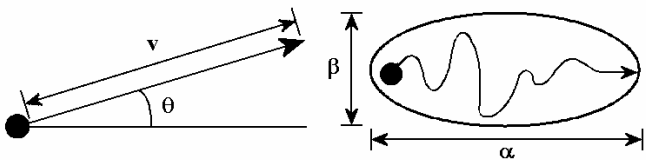

Figure 2: Diagram illustrating the four mobility parameters: velocity, direction, speed maneuverability, and breadth maneuverability.

Let A be a mobile station (the moving *agent)* and let B be a stationary (*base*) station in a network using Denning-style geo-locking for an added layer of security. In our model, the geo-locking function takes shape, time, velocity, direction, and two maneuverability parameters. The shape parameters define an ellipse as the decryption zone. An ellipse is suitable for the shape of our decryption zone because it has a length and breadth, *and* when both are equal, the ellipse becomes a circle that provides uniform coverage in all directions. (A rectangle also has a length and breadth, but when both are equal, it forms a square, with non-uniform coverage.) The time parameter specifies the period during which decryption is possible. When A is in motion, B will need to calculate a time parameter that represents a *future* time when A will actually be in the decryption zone when a geo-encrypted message arrives for decipherment at A.

Figure 2 shows the four mobility parameters that a mobile node uses to advertise its movement information. The velocity parameter, $v$, describes the recipient's speed. This is the average speed at which the recipient is expected to travel. Velocity ($v$) is determined from observing the distance traveled during a specified time unit—it is automatically calculated from recent movement, not specified by a user. The direction parameter, $\theta$, describes the direction in which the recipient is traveling and is measured as the positive angle between the positive x-axis and the velocity vector on a Cartesian coordinate system.

The first maneuverability parameter, $\alpha$, is an indication of how frequently the moving recipient might need to change speeds while traveling to the new destination (how much leeway, in terms of speed changes, that should be built into the size of the decryption zone). This speed-maneuverability parameter influences the length of the ellipse-shaped decryption zone. For travel on a commercial airliner at a fixed speed, this factor would be small, while for travel on a highway, where unexpected delays might crop up, it would probably be larger.

The second maneuverability parameter, $\beta$, defines how much the moving recipient might deviate from a straight line while traveling to a new destination (how much "wiggle room" is deemed necessary, based on an assessment of the terrain being followed). This parameter affects the breadth of the ellipse-shaped decryption zone, and we refer to it as the breadth-maneuverability parameter. On a straight highway, or in an airplane, the breadth maneuverability factor will probably be small. But in mountainous terrain, or while traveling on horseback or on a winding river, the breadth maneuverability factor will be larger. It may make more sense in the real world for a mobile station to define large maneuverability values for a single, long, winding and unpredictable move, rather than frequently computing a series of small directional and speed changes while following a crooked trail.

A mobile station must determine its own velocity and maneuverability parameters, based on its recent movement and an evaluation of the terrain in question, and communicate them to other stations for use in geo-locking messages back to the moving station.

The decryption zone only needs to be large enough for A *to extract the geo-secured decryption key* within the specified time period, not for A to decrypt the accompanying message.

In the design of our model, we make several security and reliability assumptions: that routing is secure; that authentication is assured by protocols other than geo-encryption; that the GPS hardware works flawlessly, is tamper-proof and unspoofable; that transmissions use some sort of spread spectrum method in order to counter triangulation attempts by rivals searching for our stations; and that rival electronic countermeasures do not jam our mobile stations' transmitters (presumably in laptop computers).

### B. Movement Updates

In the descriptions below, the notation $E(\{C\}, L)$ means a message with contents $C$ that are geo-encrypted to the geographical area $L$. Also, node A's current location is labeled *LA* while node B's current location is labeled *LB* in the descriptions:

One of the most important control messages in our mobility model is the *movement update message*. In order for the sender to keep track of the location and movement of the mobile receiver node, the receiver itself must accumulate information about its own ongoing movement and advertise it to the sender when necessary. Using the example where A is the mobile node and B is the stationary node, each time A's change in velocity and direction is greater than a certain threshold (discussed in Section V), A sends its current movement parameters ($v$, $\theta$, $\alpha$, and $\beta$), $LA$, and the current time $t$ to B in a movement update message:

$$E(\{v,\theta,\alpha,\beta,LA,t\},LB)$$

Based on this information, node B can predict node A's future location until A sends another movement update message to B. Because B will be estimating A's location based on this movement information, there will be errors between the estimated and actual location of A. Therefore, $LA$ and $t$ are sent in the movement update message from A to B so that B can determine A's future location knowing that A was at location $LA$ at time $t$. Similarly, B has to know A's initial location at the start of the mobile geo-encryption session (i.e., at time $t = 0$).

## IV. THE MODEL EQUATIONS

Suppose the mobile node A starts at time $t_0$ at a location whose longitude and latitude values are $LA_0$ ($X_0,Y_0$), which are assumed to be initially known to node B. This could be achieved, for example, by using the geo-encryption model in [1], or by any other means. Periodically, node A collects GPS location satellite readings $LA_t$ ($X_t$, $Y_t$) at time t with $t = t_1, t_2, t_3, \ldots$ such that $t_i = t_0 + id$ where $d$ is a fixed time unit interval whose value is arbitrary but known.

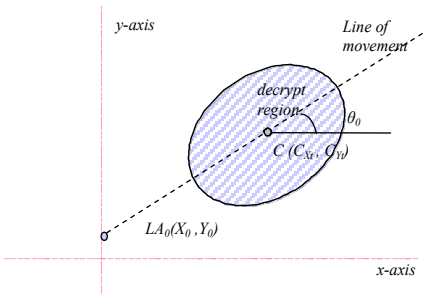

Figure 3: Movement of the decryption region depends on the initial values $V_0$, $\theta_0$ and $LA_0$.

To define the decryption region for the mobile node A, it is assumed that some initial values are available for the mobility parameters $\alpha_0$, $\beta_0$ $V_0$ and $\theta_0$ at time $t_0$. Given these initial values for the mobility parameters and $LA_0$ ($X_0,Y_0$) as the initial values for the center of the ellipse, the decryption region for node A is defined initially by substituting these values in Eq. 2.

The line of movement makes an angle $\theta_0$ with the positive direction of the latitude. As time progresses, the decryption region for node A moves along that line at a constant velocity $V_0$ (Figure 3). The movement of node A itself is arbitrary in any direction and any velocity but otherwise restricted to the decryption region at all times. The parameters of the center of the decryption region constantly change with time but not the shape. The parameters of the shape of the region remain fixed and are only allowed to change when a predetermined fixed number $n$ of time units has passed. The center ($CX_t$,$CY_t$) of the decryption region at time t is given by

$$CX_t = X_0 + (t-t_0)V\cos\theta \quad (1)$$
$$CY_t = Y_0 + (t-t_0)V\sin\theta$$

Thus, at any time node B needs only the initial parameters and the time value t to locate the center of the decryption region. On the other hand, the shape of the region is determined by the maneuverability parameters α and β as well as the movement direction θ. If we assume the region has the bivariate normal distribution with center ($CX_t$,$CY_t$) and if we adopt the 3-sigma rule [13] then the equations relating the shape parameters of the region with the maneuverability parameters are given by

$$\sigma_x = CX_t + \tfrac{1}{6}(\alpha - CX_t)\cos\theta + \tfrac{1}{6}(\beta - CY_t)\sin\theta$$
$$\sigma_y = CY_t - \tfrac{1}{6}(\alpha - CX_t)\sin\theta + \tfrac{1}{6}(\beta - CY_t)\cos\theta \quad (2)$$

Hence, at time t, the decryption region is defined by:

$$R(X_t,Y_t) = \frac{(X_t - CX_t)^2}{\sigma_X^2} + \frac{(Y_t - CY_t)^2}{\sigma_Y^2}$$
$$-2\rho\frac{(X_t - CX_t)}{\sigma_X}\frac{(Y_t - CY_t)}{\sigma_Y} \leq c \quad (3)$$

where $\rho = \cos\theta$ and c is a constant determined from values of α and β.

## V. PARAMETER ESTIMATION AND UPDATE

### A. Estimating the Mobility Parameters

Although the GPS readings were not needed to locate and determine the decryption region in the above equations, nevertheless they must be found to estimate and update the mobility parameters. The GPS readings $LA_t$ ($X_t$, $Y_t$) at time $t = t_1, t_2, t_3, \ldots$ are found and constantly used to calculate and update the velocity V, and the angle θ, of the decryption region. V and θ are the velocity and angle for movement along the line of movement. These values will be used to update the initial values $V_0$ and $\theta_0$ each time the region changes its direction or velocity. In such an event, we also update the initial values $\alpha_0$, $\beta_0$ using the last $n$ GPS readings $(X_t, Y_t)$ using the Gauss-Markov model given by

$$Z_t = \gamma Z_{t-1} + (1-\gamma)\mu + \sqrt{1-\gamma^2}\varepsilon_{t-1} \qquad t = 1, 2, 3, \cdots$$

where $0 \leq \gamma \leq 1$ is a tuning parameter representing different levels of randomness, $\mu$ is the asymptotic mean of $Z_t$, and $\varepsilon_t$ are uncorrelated stationary random Gaussian process with zero mean and unknown standard deviation $\sigma$. Notice that when $\gamma$ equals to 1 the model is identical to random walk model, and when $\gamma$ equals to zero the model is the constant velocity fluid flow model. In applications one may select the value of $\gamma$ arbitrary or could be estimated from the data. The model is described more fully in [9] and [10]. Assuming that both the velocity and the angle of the mobile node follow the model, and taking the asymptotic means of velocity and angle equal, respectively, to their initial values $V_0$ and $\theta_0$, the estimates of velocity and the angle from the $k^{th}$ period are obtained as

$$\hat{V}_k = \gamma \hat{V}_{k-1} + (1-\gamma)V_0$$
$$\hat{\theta}_k = \gamma \hat{\theta}_{k-1} + (1-\gamma)\theta_0 \qquad k = 1, 2, 3, \ldots \qquad (4)$$

where

$$\hat{V}_0 = \frac{1}{n-1}\sum_{t=1}^{n-1}\sqrt{\left(\frac{X_t - X_{t-1}}{d}\right)^2 + \left(\frac{Y_t - Y_{t-1}}{d}\right)^2}$$

$$\hat{\theta}_0 = \arctan\left\{\frac{1}{n-1}\sum_{t=1}^{n-1}\left(\frac{Y_t - Y_{t-1}}{X_t - X_{t-1}}\right)\right\} \qquad (5)$$

Notice that the $k^{th}$ period starts at time $t_{n(k-1)}$ and ends at time $t_{nk-1}$.

We obtain the formulas for estimating $\alpha$ and $\beta$ by inverting the formulas in (equations 2) to get

$$\hat{\alpha} = CX_t + 6(\hat{\sigma}_x - CX_t)\cos\hat{\theta} - 6(\hat{\sigma}_y - CY_t)\sin\hat{\theta}$$
$$\hat{\beta} = CY_t + 6(\hat{\sigma}_x - CX_t)\sin\hat{\theta} + 6(\hat{\sigma}_y - CY_t)\cos\hat{\theta} \qquad (6)$$

where

$$\hat{\sigma}_x = \frac{1}{n-1}\sum_{t=1}^{n}(X_t - \bar{X})^2, \quad \bar{X} = \sum_{t=1}^{n}X_t \Big/ n$$

$$\hat{\sigma}_y = \frac{1}{n-1}\sum_{t=1}^{n}(Y_t - \bar{Y})^2, \quad \bar{Y} = \sum_{t=1}^{n}Y_t \Big/ n \qquad (7)$$

### B. Updating the Mobility Parameters

Each time the mobility parameters are estimated, the mobile node must decide whether or not to replace the old values of the parameters with the new values and whether or not to advertise them. Typically, the old values are replaced with the new values and the updates are advertised only when they are significant i.e. when the difference between the old and the new values of a parameter exceeds some predetermined *threshold* set by the mobile node. Otherwise, the old values are kept and nothing advertised.

In addition to the mobility parameters, the initial location parameters $(X_0, Y_0)$ of the mobile node must also be updated once either or both of $V$ or $\theta$ are found significant. This is because the geo-encryption process depends on determining the center $(CX_t, CY_t)$ and, as noted from Eq.1, the recipient needs $(X_0, Y_0)$ to estimate the center. If at time $t^*$ a significant $V$ or $\theta$ is detected then not only the four mobility parameters are advertised but also the new value for $t_0$ which is estimated by $\hat{t}_0 = t^*$. Given the values of $\hat{V}, \hat{\theta},$ and $\hat{t}$ the recipient will use Equation 1 to estimate the updated initial location $(\hat{X}_0, \hat{Y}_0)$.

The smaller the threshold values for the parameters the more often the parameters are updated and advertised. So, choosing optimal threshold values is the key in optimizing the decryption zone to achieve a balance between the probability the mobile node falls inside the decryption zone against the frequency of movement updates and advertisements.

### C. Optimizing the Decryption Zone

The proposed model attempts to achieve the following goals: 1) capture the locations of other mobile stations within a given decryption region with high probability, 2) keep stations locations secret from rivals, and 3) permit the stations to be as mobile and maneuverable as possible. These three goals are in conflict and thus we need to make certain compromises in the design. For the present model, more mobility is gained at the expense of position secrecy and area of the decryption region.

In order to minimize the frequency with which nodes advertise their movements and at the same time optimize the size of the decryption zone, we propose the following:

A mobile node may fall into one of three regions as shown in Figure 4. Region 1 represents the "advertisement-free" zone, meaning that a mobile node will not advertise movement updates when they fall within this region although it constantly updates them. The size of this zone is determined by the 2-sigma rule similar to the 3-sigma rule used to define the decryption zone R(x,y) in equation 3. Hence, it is obtained by replacing 6 by 4 in equations 2. Based on the bivariate normal distribution assumption, the probability of falling in this zone is about 95% provided that the center of the region is determined according to equations 1. Regions 1 and 2 together make up the decryption zone. In region 2, the mobile node is about to leave the decrypt zone and enters the non-decrypt zone of region 3. In this zone, the node needs to transmit its mobile parameters updates if they are significant. They are declared significant when the updates in the mobility parameters exceed the parameters thresholds. Note that by restricting advertisement of updates to region 2 we effectively reduce the frequency of advertisements.

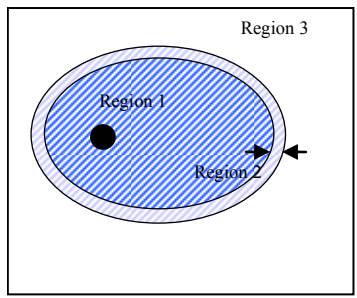

Figure 4: Regions 1, 2, and 3 for optimizing the decryption zone.

VI. PROTOCOL OVERVIEW

Our protocol builds on top of existing wireless multi-hop routing protocols, thus we will not address the routing issues of mobile multi-hop networks. We evaluated a simplified version of the Geo-encryption protocol by simulating a modified DSR protocol [11] using *ns-2* under selected scenarios (section VII). Our protocol will handle the communication of movement information between mobile nodes and the updating of this information whenever nodes move unexpectedly. It aims to allow mobile nodes to communicate their movement information accurately while at the same time reducing the overhead on the network.

When a mobile node, say MB1, wishes to communicate with another node MB2, it broadcasts a message to discover a route to MB2. When such a route exists, node MB1 will receive a route reply specifying the sequence of hops to reach node MB2. In our model, it will also need to know the position of the destination node.

This is meant to simulate a Geo-encryption model in a mobile network. Node MB1 will need to keep a table of the positions of the nodes it intends to send data to. The table will be kept current with position update messages from these nodes. Each node can obtain and send its coordinates to correspondent nodes using a position update message. The destination node MB2, upon receipt, will map its true position to the intended location of the message using the *shape parameters* (discussed in [1]). If the result of the mapping matches the location sent by MB1, the decryption will be considered successful else the message will be dropped. We will simplify this model to test if the node is within a given square centered on the received coordinates.

VII. SIMULATION MODEL

We decided to make the position update aspect of the protocol be reactive similarly to DSR routing. To simulate geo-encryption using DSR we made the following assumption:

1. Authentication is present and free: the establishment of communication between two nodes is assumed authenticated and anytime encryption fails the authenticated relationship is assumed reestablished automatically by the position update message.
2. Decryption of the message is assumed part of another layer. Our simulations are only concerned with position updates; if the message contains the right position parameters x and y to a certain tolerance +/- Tolerance, the message is considered decrypted.
3. Flow state in DSR was disabled to allow every message to carry its source route.
4. DSR control messages like route discovery messages are not subject to encryption and are not tested for it.

When a message is received by its destination node, if none of its protocol flags are set (RouteRequest, RouteReply, PositionUpdate), the node gets its own real X and Y values (GPS) and compares them to the x and y values in the packet header. If $x \in [$ Xreal $\pm$ Tolerance$]$ and $y \in [$ Yreal $\pm$ Tolerance$]$ the message is considered decrypted otherwise the message is considered not decrypted and the packet is passed to GeoHandler, a function that responds to decryption failures. GeoHandler is also called if the message is decrypted but its coordinates are not within half the tolerance ($x \in [$ Xreal $\pm$ Tolerance/**2**$]$ and $y \in [$ Yreal $\pm$ Tolerance/**2**$]$) this is meant to preempt future decryption failures.
GeoHandler constructs a message, puts the real X and Y values in the x and y fields, sets the Position Update Flag and sends the message on the reverse route that the received message arrived on.

When a message is received and it has its Position Update Flag set, it is not tested for decryption but is used to update the table entries corresponding to the source node that sent the message. This makes our Geo encryption protocol totally on demand and should only have overhead when decryption failures occur.

To evaluate the protocol we added lines to the trace file indicating the following events:
- A message was successfully decrypted.
- A message failed decryption. These would give a metrics of the protocol performance.
- A Position Update Message was sent.
- A position Update Message was received. These would allow us to gauge the overhead of the protocol on regular DSR.

## VIII. MOBILITY FILE AND SIMULATION RUNS:

We used a subset of bus routes from the Seattle area as our mobility file [12]. First we plotted the data for one of the files at our disposal and, after proper unit conversion, determined a *1500*x*1500* m area that presented dense traffic. We then selected the movements of buses within that area during a *15* minute period (simulation target time). Furthermore we excluded from that selection the buses that came to close to the edge of our area (*150* m) to avoid sudden node disappearances during the simulation. And finally, from the remaining data, we selected the *50* buses with the most number of updates in the given period (see Figure 5). From that we created an initial position file and movements file with TCL commands to include in our simulation file. We also included different pause times between movement updates to create several movement files with decreased mobility; the pause times are *10*, *25*, *50*, *75*, *100*, *200*, *400*, *650* and *900* seconds. Movements updates that exceeded *900* seconds were purged from the files, thus in effect the *900* second pause time corresponds to zero mobility when all the nodes stay at their original positions.

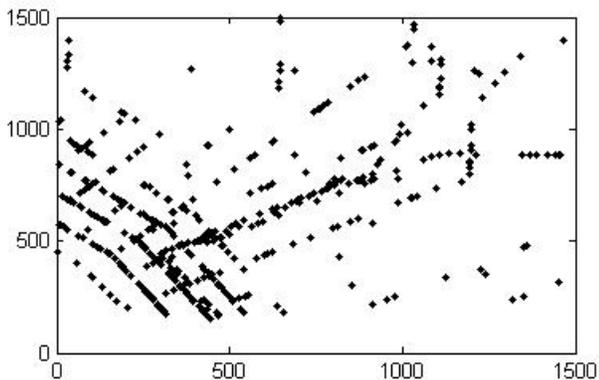

Figure 5: Final simulation data

We used *ns-2.28* for our simulations, running DSR with flow state disabled and the modifications described above.
For each one of the mobility files we did three runs with *10* sender *10* receivers, *20* sender *20* receivers and *30* sender *30* receivers. Every sender sent data at a Constant Bit Rate CBR of *4* packets per seconds with a packet size of *256* bytes.
For each run we recorded the decryption ratio and the protocol overhead.

We measured our decryption ratio as the ratio of successfully decrypted messages amongst those that were received. Thus the ratio does not reflect the delivery ratio of DSR. We measure the protocol overhead for position updates as the ratio of position update messages to the total number of data messages (CBR), decrypted or not that were received. A better measure would have been the ratio of generated position update messages to the total number of messages on the network including DSR messages as this is a modification to DSR.

## IX. SIMULATION RESULTS

With the tolerance fixed to *10* m and running for *10*, *20* and *30* senders and receivers respectively, we note that the general trend is, as expected, that the decryption ratio falls with an increase in mobility (i.e. the ratio increases with bigger pause times). See Figure 6. This is due to the fact that higher mobility means that nodes move more often away from their perceived positions at the sending nodes. As a result more messages are not decrypted.

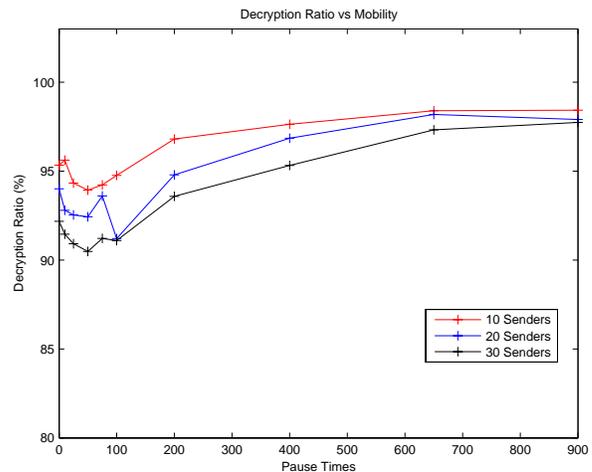

Figure 6: Decryption ratio vs. mobility for 10, 20, and 30 CBR sources

Also from Figure 6 we can see that an increase in CBR senders and receivers results in a decrease in decryption. That can be explained by the increased delay in message delivery due increase network congestion. If a message is buffered often along the way, it allows time to the destination node to move further away from its current position and thus increase the change of a decryption failure. This is confirmed by the fact that at low mobility, the gap between the ratios of the three cases is reduced. A noticeable feature of the graph is also the fast dip for pause times between *0* and *100* seconds. We have yet to explain it.

On the other hand, overhead decreases with increased pause times. See Figure 7. This behavior is typical of a protocol that is reactive to movement. If there is no movement then there is no need for movement updates. The overhead does not quite go to zero at zero mobility because in our simulation, nodes do not initially know the positions of their destinations but have to learn them by sending the default coordinates *(-1,-1)* and causing a position update message.

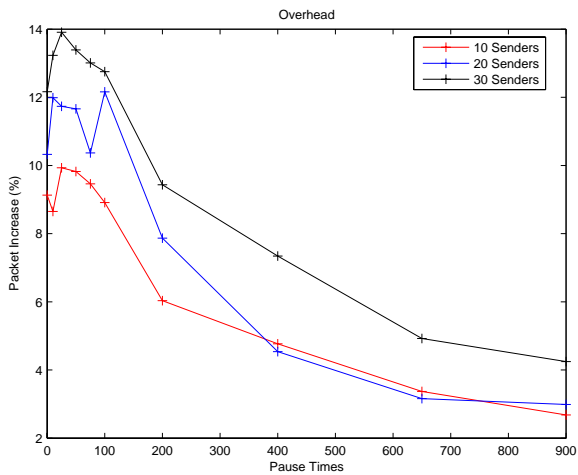

Figure 7: Protocol overhead vs. mobility for 10, 20, and 30 CBR sources

We can also see that in general, overhead increases with an increase in CBR senders and receivers. This must follow from the drop in decryption ratio discussed previously as every failed decryption or near failed (not within half tolerance) generates an update message and thus increases the overhead of the protocol. By the same token, the initial increase in overhead for small pause times compared to zero is a result of the equivalent behavior of the decryption ratio in Figure 6. We do not consider the drop of the overhead of the *20* CBR below that of *30* CBR, to be significant. We believe it to be an artifact that would disappear if we ran more simulations.

For our second series of simulations, we used our base case of *10* senders and *10* receivers to test the effect of changing the tolerance. First we compared the decryption ratio of our reference tolerance of *10* to that of a more restrictive tolerance of *3* meters. The results over the usual pause times going from *0* to *900* seconds can be seen in Figure 8.

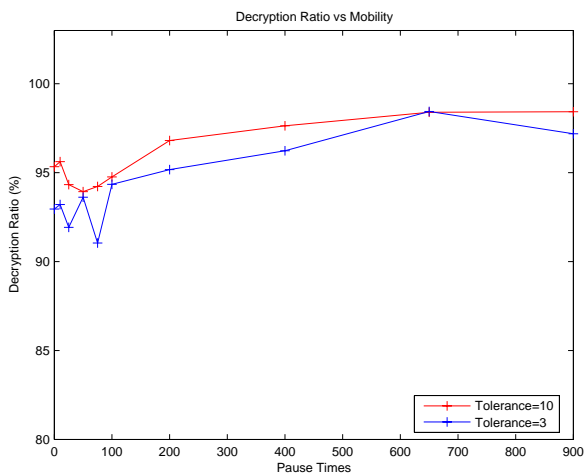

Figure 8: Decryption ratio vs. mobility for tolerance of 3 and 10

The value of the tolerance sets the size of the decryption square. Thus a more restrictive tolerance will result to a smaller square and lower decryption ratios as seen in the figure.

This trend can be further seen in Figure 9 where we show the results of increasing the tolerance through *5*, *10*, *20*, *50*, *70* and *100* meters for the maximum mobility case and with *10* CBR sources. As expected, decryption increases with more slack tolerances, that ratio would not reach *100*% though due to the need to establish the knowledge of the initial positions.

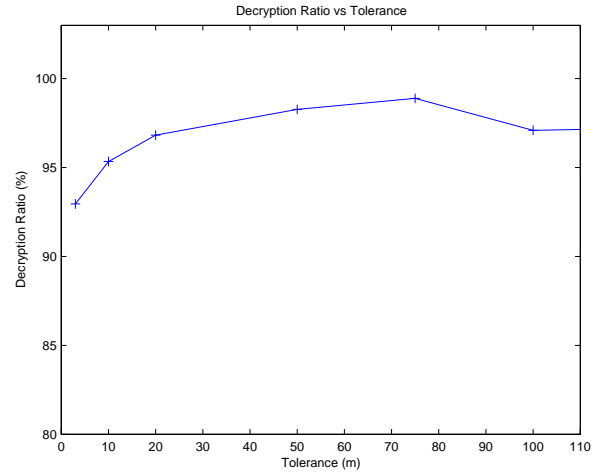

Figure 9: Decryption ratio vs. tolerance for 10 CBR sources

In a similar behavior, the overhead drops rapidly with increased tolerance down to the minimum required for the initial position messages and remains constant after a certain point. That behavior is mostly dependent on the range of motion of the buses during the simulation. If it does not exceed half of the given tolerance then no movement updates are required for the life of the simulation.

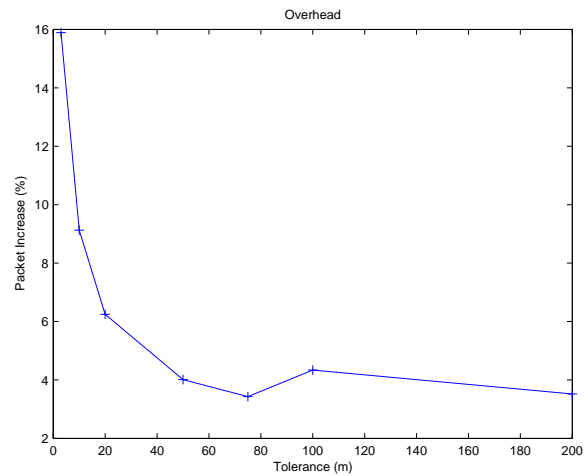

Figure 10: Protocol overhead vs. tolerance

The exponential shape can be explained by the possibility that a lot of buses have a limited range of motion and the slightest increase in tolerance is sufficient to cover their entire or most of their path and make updates unnecessary while a few buses have larger ranges of motion and will be affected only by a large increase in the tolerance.

## X. Conclusion

Using geo-encryption adds a significant layer of security to network transmissions. Mobile networks should be able to take advantage of this technique. We believe that our model serves a source for further work on location-based security for mobile networks. In our proposal, mobile nodes which stray from their advertised locations can reestablish a secure status within the network at the same time do not sacrifice the secrecy of their locations. We evaluated a simplified version of the Geo-encryption protocol by measuring the induced overhead to the network and its decryption performance by simulating a modified DSR protocol using *ns-2* under selected scenarios. Our results proved some of our expectations about decryption decline with increase in mobility and an equivalent increase in overhead. We also saw some results we did not predict such as the decrease in decryption ratio with an increase of network traffic due to increased message queuing delay.

Finally we can point certain steps to improve the performance of the protocol as next position prediction at the sender or the receiver based on history of movement, or the sending of movement parameters such as speed and direction by the receiver to the sender. And for improved security we can extend our protocol to a multi-hop encryption scheme that would require the sender to have knowledge of the position of all the forwarding nodes.

## Acknowledgement

We thank Dorothy Denning for providing us with a paper detailing her work on geo-encryption. We also thank Joe Baird and Weng Liong Low for their assistance in the preparation of an early version of this work. We are indebted to Anastasios Giannoulis and George Abi Nader for their help with the *ns-2* simulations.